\documentclass{PoS}

\usepackage{graphicx,url}

\newcommand{\td}[3]{\frac{d^{#3} #1}{d {#2}^{#3}}} 
\renewcommand{\v}[1]{\ensuremath{\mathbf{#1}}} 

\renewcommand{\bar}[1]{\ensuremath{\overline{#1}}}

\title{Gamma-rays and LHC-inspired dark matter}

\ShortTitle{Gamma-rays and LHC-inspired Dark Matter}

\author{\speaker{Geoff Beck}\\
School of Physics, University of the Witwatersrand, Private Bag 3, WITS-2050, Johannesburg, South Africa\\
E-mail: \email{geoffrey.beck@wits.ac.za}}

\author{Sergio Colafrancesco\\
School of Physics, University of the Witwatersrand, Private Bag 3, WITS-2050, Johannesburg, South Africa\\
E-mail: \email{sergio.colafrancesco@wits.ac.za}}

\abstract{The Madala hypothesis was formulated to explain anomalies in LHC data from run-1. Subsequently, these anomalies have endured into run-2 and been strengthened. This makes the analysis of the proposal highly pertinent, and, since some of the properties of the model are beyond the reach of current collider data it is also important to determine alternative means of analysis. Here, we examine the consequences of WIMPs linked to the Madala hypothesis providing a dark matter candidate and annihilating in the galactic centre of the Milky-Way and the Andromeda galaxy. These targets have been observed to have similar gamma-ray spectra by Fermi-LAT and their emissions have been widely discussed in terms of dark matter annihilation in the literature.

We show that, when the decay branchings of the hidden-sector mediator in the Madala hypothesis are assumed Higgs-like, the emissions of Andromeda and the Milky-Way are not compatible with both being produced by dark matter annihilation, apart from when a steeply contracted NFW profile is assumed for the halos or the WIMP has a mass above $1$ TeV. Additionally, similar results are displayed for a wide variety of model independent cases.}

\FullConference{5th Annual Conference on High Energy Astrophysics in Southern Africa\\
4-6 October, 2017\\
University of the Witwatersrand (Wits), South Africa}

\begin{document}

\section{Introduction}
The Madala hypothesis was proposed in 2015 to account for anomalies appearing in the run-1 data from LHC~\cite{madala1,madala2} in both the ATLAS~\cite{atlas-docs} and CMS experiments~\cite{cms-docs}. It introduces a heavy Higgs-like `Madala' boson, a Dark Matter (DM) candidate $\chi$ and a hidden sector mediator boson $S$ of similar mass to the Higgs particle. In the second run the motivating anomalies have both remained and grown more significant~\cite{madala3,madala4,madala5}. However, the properties of the scalar mediator $S$ are only lightly constrained by existing collider data~\cite{madala2,madala3,madala4,madala5}. Previous efforts by the authors of this work~\cite{gsmadala} have examined the potential constraints on the Madala hypothesis from the spectrum of the Milky-Way Galactic Centre (GC) excess observed by Fermi-LAT~\cite{fermi-docs,fermigc2015}, as well as with radio data from the Coma cluster~\cite{gsmadala2}. The previous work in the GC was conducted under the assumption of a contracted Navarro-Frenk-White (NFW) profile for the GC, but evidence exists to argue for a considerably flatter profile~\cite{nesti2013} (similarly in Andromeda (M31) and galaxies in general~\cite{banerjee,rodrigues2017}). Therefore, we will generalise this analysis to include recent Andromeda spectral data from Fermi-LAT~\cite{fermim312017} and to accommodate more general halo profiles (other DM work has been done with Andromeda spectra by the authors in \cite{gsm31}).

The approach we take in this work is to predict DM annihilation emissions from both the GC and M31 in the gamma-ray band. These will be compared to existing Fermi-LAT data~\cite{fermigc2015,fermim312017} to derive two things, the first being the best-fit DM models for the GC excess, and the second being the $3\sigma$ confidence level exclusion limits from M31. With these in hand we can determine whether the gamma-ray spectra for the GC and M31 are compatible with being produced by the same DM models. The chosen compatibility requirement will be that the GC best-fit cross-section is not excluded to $3\sigma$ confidence level by the limits from M31. In particular, we will focus on models associated with the Madala hypothesis and we will determine if the assumption of Higgs-like Standard Model decay branching~\cite{smhiggs} for the mediator boson $S$ are compatible with a DM origin for gamma-ray observations of the GC and M31. We study the parameter space of the mass of $S$ between $130$ and $150$ GeV in accordance with LHC run-2 favoured masses~\cite{madala5}.

We find that, for general WIMP scenarios with an NFW halo profile, DM models which annihilate via heavy leptons with $m_{\chi} > 10^2$ GeV, or via muons, electrons, and photons with $m_{\chi} \approx 10$ TeV are compatible with both galactic spectra. However, almost all annihilation channels and masses are compatible when a contracted NFW halo profile is assumed. The exceptions being annihilation via quarks or weak bosons when $m_{\chi} < 100$ GeV. For the Madala scenario with NFW halos, Higgs-like $S$ decay branchings are only compatible with both gamma-ray spectra when the WIMP exceeds a mass of $1$ TeV, this being a region also unconstrained by previous work on indirect astrophysical probes of the hypothesis~\cite{gsmadala2}. Importantly, the favoured region of parameter space for the galactic centre excess~\cite{calore2014,daylan2016} is the most strongly excluded when the halo profiles are shallower than the contracted NFW case. This demonstrates how sensitive these DM explanations are to details of the DM density distribution.

This paper is structured as follows: in section~\ref{sec:ann} we detail DM annihilation models, in sec.~\ref{sec:emm} the gamma-ray emission formalism is outlined, while the results are discussed in sec.~\ref{sec:results} and conclusions are drawn in sec.~\ref{sec:conc}.

\section{Dark Matter Annihilation}
\label{sec:ann}
The source function for gamma-rays with energy $E$ from a $\chi\chi$ annihilation (and subsequent $S$ decay when considering the Madala hypothesis) is taken to be
\begin{equation}
Q_\gamma (r,E) = \langle \sigma V\rangle \sum\limits_{f}^{} \td{N^f_\gamma}{E}{} B_f \left(\frac{\rho_{\chi}(r)}{m_{\chi}}\right)^2 \; ,
\end{equation}
where $r$ is distance from the halo centre, $\langle \sigma V \rangle$ is the non-relativistic velocity-averaged annihilation cross-section, $f$ labels the annihilation channel intermediate state with a branching fraction $B_f$ and differential gamma-ray yield $\td{N^f_\gamma}{E}{}$, $\rho_{\chi}(r)$ is the radial density profile of $\chi$ particles in the halo, and $m_{\chi}$ is the $\chi$ mass. The $f$ channels used will be quarks $q\bar{q}$, electron-positron $e^+e^-$, muons $\mu^+ \mu^-$, $\tau$-leptons $\tau^+\tau^-$, $W$ bosons $W^+W^-$, $Z$ bosons $ZZ$, and photons $\gamma\gamma$. When considering the Madala scenario the cross-section will represent an effective annihilation from $\chi\chi \to S \to SM$, as $S$ could well decay back to $\chi$ particles.

The yield functions $\td{N^f_i}{E}{}$ are taken from \cite{ppdmcb1,ppdmcb2} for all channels (with electro-weak corrections), however, when $m_{\chi} < (m_Z , \, m_W)$ the model independent formulation within the micrOMEGAs package~\cite{micromegas1,micromegas2} is used instead for the $ZZ$ and $W^+W^-$ channels.

\section{Gamma-ray Emission}
\label{sec:emm}
For the DM-induced $\gamma$-ray production, the resulting flux calculation takes the form
\begin{equation}
S_{\gamma} (\nu,z) = \int_0^r d^3r^{\prime} \, \frac{Q_{\gamma}(\nu,z,r)}{4\pi D_L^2} \; ,
\end{equation}
with $Q_{\gamma}(\nu,z,r)$ being the source function for frequency $\nu$ and position $r$ within the given DM halo at redshift $z$, and $D_L$ is the luminosity distance to the halo.
The integration over the source function $Q$ will be summarised in the astrophysical J-factor of the target halo:
\begin{equation}
J (\Delta \Omega, l) = \int_{\Delta \Omega}\int_{l} \rho^2 (\v{r}) dl^{\prime}d\Omega^{\prime} \; , \label{eq:jfactor}
\end{equation}
with $\rho (r)$ being the halo density profile, the integral being extended over the line of sight $l$, and $\Delta \Omega$ is the observed solid angle.
The flux can then be written as
\begin{equation}
S_{\gamma} (\nu,z) = \langle \sigma V\rangle \sum\limits_{f}^{} \td{N^f_i}{E}{} B_f J(\Delta \Omega,l) \; .
\end{equation}

To calculate $J$-factors we will consider generalised NFW halo profiles for both M31 and the GC
\begin{equation}
\rho(r)=\frac{\rho_s}{\left(\frac{r}{r_s}\right)^\gamma\left(1+\frac{r}{r_s}\right)^{3-\gamma}} \; ,
\label{eq:nfwcon}
\end{equation}
with $\rho_s$ normalising the profile to the halo mass, $r_s$ being the halo scale radius, $\gamma = 1.3$ for a contracted NFW halo (for comparison to GC best-fit DM models~\cite{daylan2016}), and $\gamma = 1$ yielding vanilla NFW~\cite{nfw1996}.
For the GC we use the value of $J$ found in \cite{fermigc2015} giving the value in the observed region as $2\times 10^{22}$ GeV$^2$ cm$^{-5}$ for a NFW profile, and $J \sim 4 \times 10^{23}$ GeV$^2$ cm$^{-5}$ for a contracted NFW profile following the calculations of \cite{daylan2016}. While, in M31, we use the value from \cite{fermim312017} of $J \sim 8 \times 10^{18}$ GeV$^2$ cm$^{-5}$ for the region observed with an NFW halo ($\sim 1 \times 10^{20}$ will be used for the case of a contracted NFW halo following the same method as above). Boosting of annihilation rates from dense substructure within the target halos is not considered here.

\section{Results}
\label{sec:results}
The results displayed here make use of two sets of limits on $\langle \sigma V \rangle$. The first is obtained from M31 Fermi-LAT data by searching for the smallest values of $\langle \sigma V \rangle$ that are excluded at $3\sigma$ by the data. We do this by comparing predicted DM emissions within the same region of M31 as observed in \cite{fermim312017} to the data obtained in the aforementioned work. In the case of the GC we similarly compare predicted DM emissions within the same region is observed in \cite{fermigc2015} and compare these to the spectra found by Fermi-LAT to produce values of $\langle \sigma V \rangle$ that best-fit the data. This approach allows DM emissions to account only for some part of the observed spectra with this contribution being limited by existing data. As we do not account for other spectral contributions the limits produced are likely conservative but still provide robust insights. We stress that, although we could determine a best-fit as well, we use an upper limit in the case of M31 in order to check whether models that fit the GC case are in fact ruled out by the M31 case, as this provides a more stringent test of compatibility between the two sources.  

\begin{figure}[htbp]
\begin{center}
\resizebox{0.8\hsize}{!}{\includegraphics{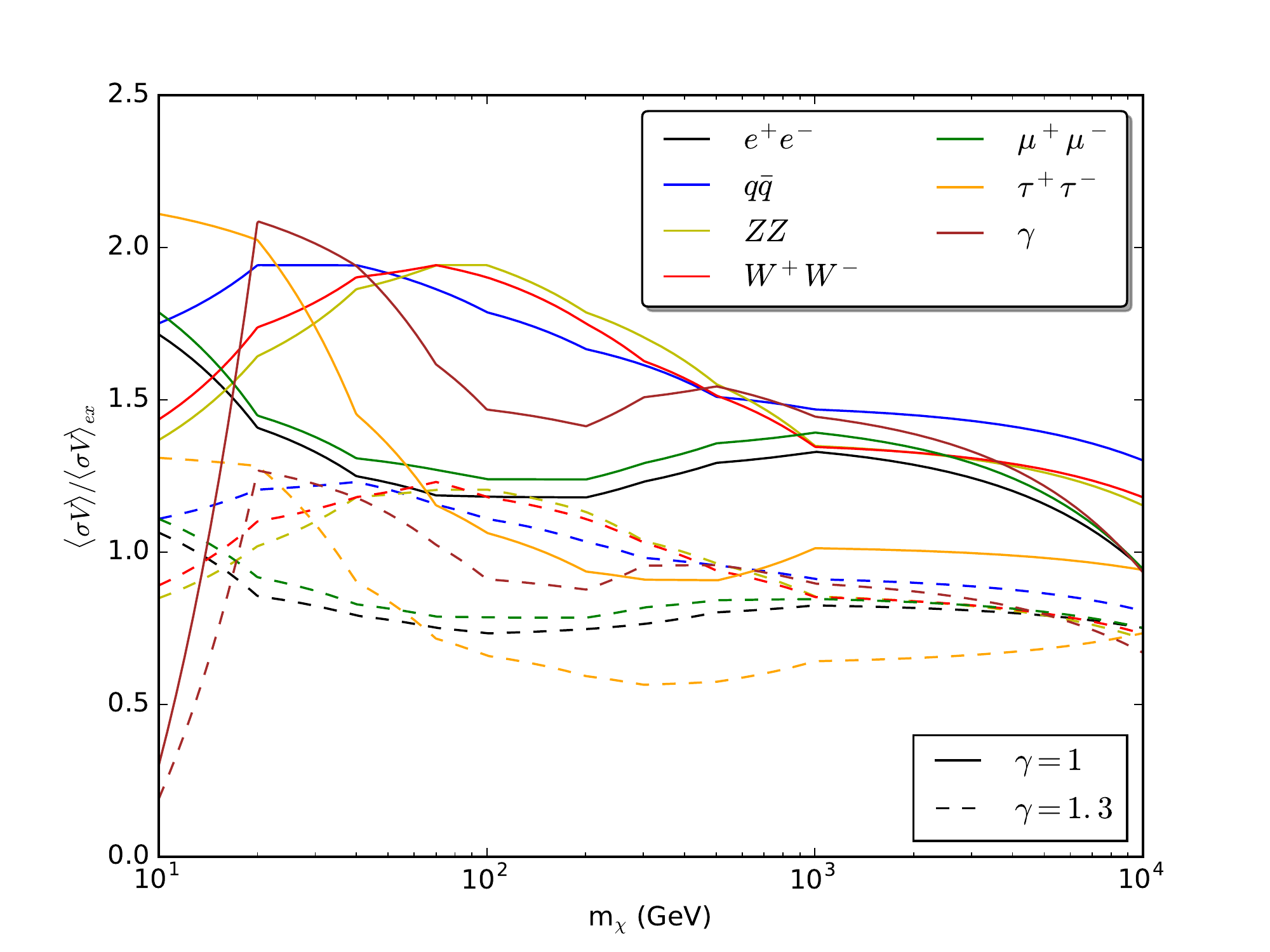}}
\end{center}
\caption{Ratio of best-fit GC cross-sections to the exclusion limits at $3\sigma$ confidence level from M31. Here a range of annihilation channels are displayed individually. The $\gamma$ parameter is $1$ for plain NFW (solid lines) and $1.3$ for the contracted halo profile (dashed lines). }
\label{fig:bestfit}
\end{figure}

Figure~\ref{fig:bestfit} displays the compatibility for each WIMP annihilation channel considered individually. This is taken to be the ratio of the best-fit cross-section for the GC to the $3\sigma$ confidence level exclusion from M31. With an NFW profile, all the channels except $\tau$-leptons have the best-fit GC models excluded by M31 spectral data below WIMP masses of 10 TeV. The aforementioned heavy lepton channel reaches compatibility around $100$ GeV. Approaching $10$ TeV WIMP masses the electron, muon, and direct photon channels become compatible with both spectra even when an NFW profile is adopted. When a contracted NFW profile is assumed all the channels are compatible above $100$ GeV, but several channels show tension between the spectra below this mass threshold, particularly quarks and weak bosons. This is significant since the models proposed for the GC excess lie in the region below $100$ GeV WIMP masses~\cite{calore2014,daylan2016}.

\begin{figure}[htbp]
\begin{center}
\resizebox{0.8\hsize}{!}{\includegraphics{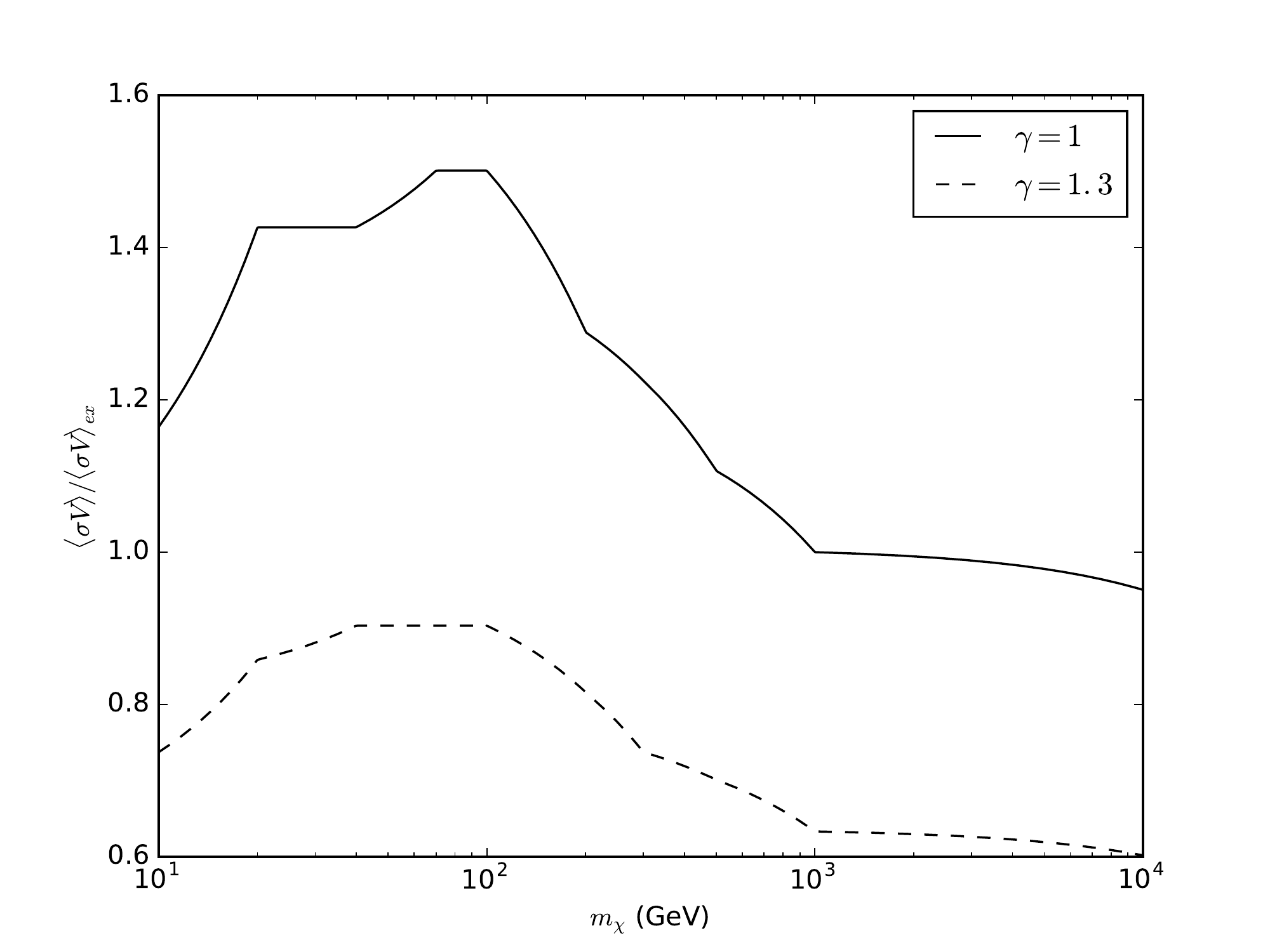}}
\end{center}
\caption{Ratio of best-fit GC cross-sections to the exclusion limits at $3\sigma$ confidence level from M31. In this case annihilation channels are combined assuming $\chi\chi \to S$ with $m_{S} = 130$ GeV using Higgs-like data~\cite{smhiggs}. The $\gamma$ parameter is $1$ for plain NFW (solid lines) and $1.3$ for the contracted halo profile (dashed lines). }
\label{fig:madalafit}
\end{figure}

In Figure~\ref{fig:madalafit} we see the compatibility between the GC best-fit and M31 exclusions for the Madala scenario where the decay branchings of $S$ are Higgs-like with $m_{S} = 130$ GeV. In this case we can see the contracted profile allows compatibility for all WIMP masses shown. However, with an NFW halo profile, all WIMP masses below $1$ TeV are incompatible, with this peaking between $10$ and $100$ GeV in the aforementioned region for the GC excess. These results do not qualitatively change when the mass of $S$ is extended out to 150 GeV, which is the maximum extent of the currently favoured parameter space~\cite{madala5}.

\section{Conclusions}
\label{sec:conc}

In this work we have shown that for WIMP masses favoured by the GC gamma-ray excess~\cite{calore2014,daylan2016}, and Higgs-like $S$ decay branchings, the Madala hypothesis DM candidate cannot account for both the M31 and GC gamma-ray spectra when the density profile of the DM halos is assumed NFW or shallower. This is true across the possible mass range of the $S$ boson~\cite{madala5}. However, larger wimp masses $m_{\chi} > 1$ TeV can still be compatible with both galactic spectra and are unconstrained by other indirect methods~\cite{gsmadala2}.

For general WIMP models we demonstrate that all annihilation channels are compatible with both galactic spectra, across the studied WIMP mass range $1$ - $10^4$ GeV, when a contracted NFW density profile is assumed for the halos, apart from via quarks and weak bosons below $100$ GeV WIMP masses. However, when a shallower profile is considered, the $\tau$ lepton channel is compatible when $m_{\chi} > 100$ GeV, while annihilation via photons and light leptons is only compatible when $m_{\chi} \approx 10$ TeV. Other channels are incompatible with producing the two spectra across the mass range when a shallower halo density profile is assumed. 

This compatibility study, using cross-section ratios, is robust as it minimises the systematic errors incurred when estimating galactic $J$-factors. Additionally, these results are especially relevant given evidence favouring cored halo density profiles in galaxies~\cite{banerjee,nesti2013,rodrigues2017}.

\section*{Acknowledgements}
This work is based on the research supported by the South African Research Chairs Initiative of the Department of Science and Technology and National Research Foundation of South Africa (Grant No 77948).
G.B. acknowledges support from a post-doctoral grant through the same initiative and institutions.
We acknowledge useful and stimulating discussions with B. Mellado on the Madala hypothesis proposed by members of the Wits-ATLAS group to account for several anomalies in both ATLAS and CMS data at the LHC.

\end{document}